# Comparative computational study of lithium and sodium insertion in van der Waals and covalent tetracyanoethylene (TCNE) -based crystals as promising materials for organic lithium and sodium ion batteries

Yingqian Chen and Sergei Manzhos[*]

Department of Mechanical Engineering, National University of Singapore, Block EA #07-08, 9 Engineering Drive 1, Singapore 117575, Singapore

*E-mail: mpemanzh@nus.edu.sg; Tel: +65 6516 4605

**Abstract**

We present a comparative ab initio study of Li and Na insertion in molecular (van der Waals) crystals of TCNE (tetracyanoethylene) as well as in covalent Li/Na-TCNE crystals. We confirm the structure of previously synthesized (covalent) Li-TCNE crystal as well as predict the existence of its Na-TCNE analogue. In the molecular/covalent TCNE crystals, insertion sites are identified with the binding energy of Li and Na up to 2.7/1.8 and 2.6/1.8 eV stronger than Li and Na cohesive energy, respectively, in dilute concentrations. Up to 5.5/2.5 and 3/2 Li and Na atoms per TCNE unit can be inserted in the molecular/covalent crystals, respectively, while preserving the structure, with maximum voltages, respectively, 3.5/2.2 and 3.3/2.7 V. Significantly, up to capacity of 418 mAh g$^{-1}$ for both Li and Na in the molecular crystal and 198 mAh g$^{-1}$ for Li and 177 mAh g$^{-1}$ for Na in the covalent crystal, the insertion of Li and Na would not lead to reactions with common electrolytes. Tetracyanoethylene-based molecular and covalent crystals could therefore become efficient organic cathode *and* anode materials for Li *and* Na ion batteries.

## 1. Introduction

Development of novel electrochemical storage technologies is important for a range of applications, not the least for grid storage including peak shifting and price arbitrage [1, 2]. This is critical to achieve widespread use of renewable but intermittent sources of electricity (such as wind and solar) as well as a larger share of (all-)electric vehicles which are expected to have longer range and better efficiency in the future [3]. Lithium ion batteries provide today the highest energy density, cycle rate, and cycle life among commercial batteries [4]. However, further improvements in performance as well as in sustainability are required [5]. Especially for efficient grid storage, rapid charge-discharge (within minutes) is needed [1, 6], which is beyond that of present commercial Li ion batteries [5]. In addition, some expensive or poisonous components like Co [7] are used in present electrode materials of Li ion batteries. Further, lithium deposits are geographically concentrated and may be insufficient to use Li ion batteries on a large scale [8-10]; Li might become too costly with the growing battery market. On the other hand, sodium is abundant and cheap, relatively light, and has qualitatively similar valence shell chemistry as lithium. Sodium ion batteries are one promising candidate technology for bulk electrochemical storage. However, compared to materials for Li storage, it seems to be more difficult to design inorganic electrode materials with suitable thermodynamics and kinetics for Na storage [11-14]; this is due, in particular to the larger atomic radius of $Na^+$.

Organic electrodes for rechargeable batteries have been receiving increasing attention in recent years, as they are a way to achieve simultaneously high rate (high power) and environment-friendly batteries. They can, for example, be made from common feedstock such as biomass [15-17]. In recent years, significant progress in the development of organic batteries has been made [15]. Capacities of up to 900 mAh $g^{-1}$ (i.e. competitive with inorganic electrode materials) and rates of up to 1000C (unprecedented for inorganic electrodes) have been reported [15]. Moreover, organic electrodes are also promising for post-Li storage [18, 19], which will have to be developed to make massive deployment of electrochemical batteries feasible [20].

A number of experimental works on different classes of potential organic electrode materials for Li ion batteries [15, 18, 21-24] have been done in recent years. Computational studies are, on the other hand, very few [25, 26]. Modeling is, however, important to screen for potential new electrode materials, to explain the mechanism of operation of known electrode materials, and ultimately to guide experimental design towards more performant materials. For Na storage, several organic materials have been proposed, including carboxylate and terephthalate based materials [18, 22]. One promising class of organic electrode materials is tetracyanides. Cathodes for Li ion batteries made of crystalline tetracyanoquinodimethane (TCNQ) [21] were reported to achieve a relatively high capacity exceeding 200 mAh g$^{-1}$ with excellent cyclability. The reported lithiation mechanism involves coordination of Li to CN moieties; that is to say, the aromatic ring would not contribute to the capacity. Further, the voltages reported for this material are 2.5~3.2 V which is not optimal for either cathodic (where voltages $\gtrsim$ 4 V are desired) or anodic (where voltages closer to 0 are desired) operation. Therefore, here, we consider a smaller tetracyanide molecule without the (potentially dead-weight) aromatic ring, tetracyanoethylene (TCNE) [27] as a potential new organic electrode material which is expected to have high specific capacity and voltages more suited for anodic operation [26, 28, 29].

TCNE is a lighter molecule which is expected to attach up to 4 Li/Na atoms [27], which would result in a theoretical specific capacity of about 840 mAh/g$_{(TCNE)}$. In Ref. [26], we studied ab initio lithium and sodium attachment to TCNE *molecules*, both free and attached to (doped) graphene. We predicted that up to four (five) Li and Na atoms can be stored on free (adsorbed) TCNE with binding energies stronger than cohesive energies of the Li and Na metals. Interestingly, there was no significant difference either in specific capacity (per unit mass of material excluding Li/Na) nor in predicted voltage between Li and Na storage. In contrast, for many inorganic electrode materials, Na insertion is thermodynamically inhibited compared with Li insertion [11-14], which makes organic molecules very promising for post-Li storage in general. TCNE, therefore, is a promising candidate molecule for organic Li and Na ion batteries. However, TCNE is stable and easily available under normal conditions in a crystalline form. It is known to form two types of molecular (vdW-bound) crystals: a cubic and a monoclinic phase [30]. Contrary to the monoclinic phase, the cubic phase is not stable and transforms into the monoclinic when the temperature is higher than 320 K, which then remains monoclinic upon cooling. Therefore, to understand practical potential of TCNE as organic electrode, we here study Li and Na interaction with *crystalline* TCNE, specifically, the monoclinic phase (left panel in Fig. 1). Recently, TCNE has been reported to form *covalent* crystals with Li of stoichiometry Li-TCNE [27] which presents well-defined channels that could be suitable for Li storage and transport (top left panel in Fig. 4). The potential of this material as organic electrode has not been studied. Also, the existence and potential for Na storage of its putative Na-TCNE analogue remain unknown.

Here, we present a comparative dispersion-corrected density functional theory (DFT) computational study of the possibilities of Li and Na storage in tetracyanoethylene-based molecular (vdW) and covalent crystals. Specifically, we confirm the previously reported XRD structure of Li-TCNE [27] and predict the existence of a covalent Na-TCNE crystal. We identify Li and Na insertion sites and compare the energetics and voltages as well as the theoretical capacities of Li vs. Na storage in vdW vs. covalent crystals.

## 2. Methods

Crystalline structures were optimized with DFT [31] using the SIESTA code [32]. The PBE exchange-correlation functional [33] and a DZP (double-ζ polarized) basis set were used. The basis set was optimized to reproduce the cohesive energies of C, Li, Na and N [34-36]. Specifically, the cohesive energy of Li metal is $E_{coh}^{Li}$ = -1.67 eV and $E_{coh}^{Na}$ = -1.14 eV computed with these basis sets is very accurate [34, 35] and can be relied upon to compute the voltages. Geometries were optimized until forces on all atoms were below 0.02 eV/Å. Simulation cell vectors were optimized until stresses were below 0.1 GPa. A cutoff of 200 Ry was used for the Fourier expansion of the density, and a *bcc* type oversampling of the Fourier grid was used to minimize the eggbox effect. Smearing equivalent to an electronic temperature of 500 K was used to speed up convergence. To find Li/Na insertion sites, a periodic supercell of size ~11×11×12 Å (corresponding 2×2×1 unit cells) was used for the covalent crystals, with 8 TCNE-Li/Na molecules per supercell. The Brillouin zone was sampled with a 2×2×2 grid of Monkhorst-Pack points [37]. For molecular (vdW) crystals, insertion sites were computed in a periodic supercell of size 14×12×14 Å (corresponding to 2×2×2 unit cells) with 16 TCNE molecules per supercell. The Brillouin zone was sampled at the Γ point. Spin polarization was used in all calculations. Stability of the crystal structures and of insertion sites was confirmed by quenched molecular dynamics (MD) calculations performed following geometry optimization, whereby the structures were relaxed until forces on all atoms were below 0.015 eV/Å. No appreciable geometry or energy changes were detected.

For modeling of insertion of multiple Li/Na atoms into the covalent crystals, the unit cell (2 TCNE-Li/Na units) was used with a 9×9×5 Monkhorst-Pack point grid. For vdW crystals, the unit cell (2 TCNE molecules) was used for insertion of multiple Li/Na atoms with a 4x4x4 Monkhorst-Pack point grid. Calculations for insertion of 0.5 Li/Na atoms per TCNE were also computed in the supercell described above (i.e. corresponding 2×2×1 unit cells). For each specific capacity (no. of inserted Li/Na atoms), the configuration with the strongest binding was also rechecked by quenched MD to confirm its stability, as described above.

Charge transfer between Li/Na atoms and TCNE-based crystals was analyzed using Mulliken charges as well as Voronoi charges [38] (which are defined in a basis-set independent way). Dispersion forces between TCNE units, which are important for the modeling of vdW-bound crystals, were modelled with the scheme of Grimme [39] (DFT-D2) with parameters taken from Ref. [39].

The binding energy ($E_b$) per Li/Na atom was computed as

$$E_b = \frac{E_{nX/sys} - E_{sys} - nE_X}{n} \qquad (1)$$

where $E_{nX/sys}$ is the total energy of $n$ X atoms inserted into *sys*, where *sys* is the molecular or covalent crystal and $X$ = Li or Na; $E_{sys}$ is the total energy of *sys*, and $E_X$ is the total energy of an $X$ atom in a vacuum box (of the same size as the supercell). A negative value of $E_b$ therefore corresponds to thermodynamically favored insertion.

The average voltage between concentrations $x$ and $x_0$ of Li or Na is computed using the following equation [40, 41]:

$$V = -\frac{E_x - E_{x_0} - (x-x_0)E_{X(bcc)}}{q(x-x_0)}, \qquad (2)$$

where $E_{x(0)}$ is the energy of the host material with Li/Na concentration $x_{(0)}$, $E_{X(bcc)}$ is the energy of atom $X$=Li/Na in its *bcc* structure, and $q$ is the net charge of the $X$ ions ($q$ = +1 $e$).

## 3. Results

*3.1 Li and Na insertion into vdW crystals*

The crystal structure of the TCNE molecular crystal was taken from Ref. [30] and was optimized with the present setup (the first configuration in Fig. 1). The optimized lattice parameters of this crystal are $a$ = 7.38 Å, $b$ = 5.87 Å and $c$ = 6.70 Å; $\alpha$ = 90.00°, $\beta$ = 97.58° and $\gamma$ = 90.00°. These parameters can be compared with experimental results that range: $a$ = 7.48-7.51 Å, $b$ = 6.20-6.21 Å and $c$ = 6.99-7.00 Å; $\alpha$ = 90.00°, $\beta$ = 97.10-97.35° and $\gamma$ = 90.00° [30, 42-45]. The agreement can be considered good for vdW systems [24, 46]. The cohesive energy of this crystal is -1.39 eV per TCNE molecule.

We performed a search for possible insertion sites by inserting Li/Na atoms in many possible locations within the supercell and performing optimization. Four stable non-equivalent insertion sites were found in the molecular crystal for both Li and Na insertions (Fig. 1), with $E_b$ stronger than $E_{coh}$ of the Li and Na metals. The lowest energy site for Li/Na has $E_b$= -4.36/-3.71 eV which is 2.69 and 2.57 eV stronger than the $E_{coh}$ of Li and Na, respectively. The binding energies vs. $E_{coh}$ of Li and Na are shown in the leftmost red empty squares of Fig. 7, top panels. This means that insertion of dilute concentrations of Li and Na is thermodynamically favored at potentials about 2.7 and 2.6 V vs Li/Li$^+$ and Na/Na$^+$, respectively, making the molecular TCNE crystal a potential cathode materials.

We then studied insertion of multiple Li and Na atoms into the molecular TCNE crystal. Multiple configurations were tried. We found that up to 5.5/3 Li/Na atoms per TCNE unit can be inserted while preserving the crystal structure. This corresponds to 1151/628 mAh/g capacity for Li/Na. The binding energies vs. $E_{coh}$ of Li and Na are shown in Fig. 7, top panels. For each no. of inserted atoms, the configuration with the strongest $E_b$ is shown in Fig. 2 and 3. For Li insertion, at higher concentrations, there are significant distortions of the TCNE molecular structure, so that a four-membered cyclic structure is observed. These, however, relax back to the original TCNE structure after removal of Li.

As can be seen in Fig. 7, top panels, the binding of Li/Na atom in the molecular crystal is strongest when 1/0.5 Li/Na atom per TCNE molecule is inserted in the crystal (corresponding to a specific capacity of 209/105 mAh g$^{-1}$). Beyond this point, the binding of Li/Na atoms would be weakened. This means that during the insertion of up to 1/0.5 Li/Na atom per TCNE molecule, Li and Na atoms will concentrate into zones with these concentrations, i.e. a separation of lithiated/sodiated and non-lithiated/sodiated phases is expected at constant voltage of 3.54/3.31 V vs Li/Na bulk. The expected voltage profiles for Li and Na insertion are shown in Fig. 7, bottom panels, taking into account expected phase segregation. The profiles confirm that molecular TCNE crystals can be used as organic cathodes of Li and Na ion batteries. Significantly, up to the capacity of 418 mAh g$^{-1}$ for both Li and Na insertion, the voltage is within the electrochemical stability window of common liquid organic electrolytes, such as LiPF$_6$ in EC:DEC between 1.3 V and 4.5V [47] and NaClO$_4$ in EC:DMC between 1.2 V [48] and 4.5 V [49] (vs. Li/Li$^+$ and Na/Na$^+$, respectively). Insertion of Li and Na in TCNE up to these capacities would therefore not lead to reactions with the electrolyte and is expected to be safe. The voltage drops to zero at 1151 mAh g$^{-1}$ for Li and 418 mAh g$^{-1}$ for Na.

The average charge donation of Li/Na in the molecular crystal was also computed. For the configurations shown in Fig. 2 and 3, up to 0.70/0.84 |$e$| (Mulliken) and 0.28/0.30 |$e$| (Voronoi) per Li/Na atom are donated to TCNE molecules, even at the largest Li/Na concentrations. This is in contrast to the Li/Na attachment to a single TCNE molecule, where there was a clear and matching weakening trends of $E_b$ and of average charge donation with the number of attached Li/Na atoms [26, 28, 29].

*3.2 Li and Na insertion into covalent Li/Na-TCNE crystals*

We also performed ab initio optimization of the covalent Li-TCNE crystal structure (top left panel of Fig. 4) and its Na analogue (top right panel of Fig. 4), both of which are stable configurations. The unit cell of the covalent Li-TCNE crystal has a structure with lattice parameters $a$ = 5.42 Å, $b$ = 5.56 Å and $c$ = 11.93 Å; $\alpha$ = 90.00°, $\beta$ = 110.12° and $\gamma$ = 90.00°. The volume is just 1% larger than that of the crystal structure reported in an experimental study [27] ($a$ = 5.43 Å, $b$ = 5.41 Å and $c$ = 11.91 Å; $\alpha$ = 90.00°, $\beta$ = 107.70° and $\gamma$ = 90.00°). After replacing Li with Na atoms, we have found a stable structure of the Na-TCNE analogue, with lattice parameters $a$ = 5.92 Å, $b$ = 5.87 Å and $c$ = 12.43 Å; $\alpha$ = 90.00°, $\beta$ = 111.40° and $\gamma$ = 90.00°. The formation energy of Li/Na-TCNE crystals is -4.82/-4.38 eV per formula unit (vs. TCNE molecules and *bcc* Li/Na, negative sign means favorable formation).

Similarly to the molecular crystal, multiple possible insertion sites were tried, and four stable non-equivalent insertion sites were found in these covalent crystals (the bottom panels in Fig. 4), with $E_b$ stronger than $E_{coh}$ of the Li and Na metals. The lowest energy site for Li/Na has $E_b$= -3.45/-2.89 eV which is 1.78 and 1.75 eV stronger than the $E_{coh}$ of Li and Na, respectively. The binding energies vs. $E_{coh}$ of Li and Na are shown in the leftmost black filled squares of Fig. 7, top panels. This means that insertion of dilute concentrations of Li and Na is thermodynamically favored at potentials about 1.8 V vs Li/Li$^+$ and Na/Na$^+$, respectively, making Li/Na-TCNE a potential anode materials.

Insertion of multiple atoms into the covalent crystals with many possible configurations was then modeled. Up to 2.5/2 Li/Na atoms per TCNE unit can be inserted in the covalent crystal which corresponds to 495/355 mAh g$^{-1}$ capacity, respectively, while preserving the crystal structure. The binding energies vs. $E_{coh}$ of Li and Na are shown in Fig. 7, top panels. For each no. of inserted atoms, the configuration with the strongest $E_b$ is shown in Fig. 5 and 6. Comparing Fig. 2 and 3 with Fig. 5 and 6, we can find that the structures of covalent crystals are much less distorted by the insertion than that of the molecular crystal, which, as expected, shows that the covalently bound Li-TCNE framework is more stable than the vdW-bound framework of the molecular crystal. Furthermore, by comparing the vacancy formation energy, which is 5.4/4.6 eV for the extraction of a Li/Na atom from the structures of shown in the top panel of Fig. 4 into vacuum to the insertion energy $E_b$, which peaks at -3.83/-3.80 eV for Li/Na insertion, we also confirmed that Li/Na-TCNE crystals are stable under Li/Na insertion/extraction.

The strongest binding energy per Li/Na atom peaks at the concentration of 1/0.5 Li/Na atoms per Li/Na-TCNE unit, which would correspond to specific capacity of 198/89 mAh g$^{-1}$ for Li/Na. Up to this capacity, therefore, a separation of lithiated/sodiated and non-lithiated/sodiated regions is expected to occur at a constant voltage of 2.16/2.67 V vs. Li/Na bulk. The computed voltage profile is shown in Fig. 7, bottom panels, taking into account expected phase segregation.

The voltage drops to zero after 397/177 mAh g$^{-1}$ for Li/Na insertion, and remains above the electrolyte reduction potential up to 198/177 mAh g$^{-1}$ for Li/Na. Therefore, up to these capacities the covalent Li/Na-TCNE crystalline electrode will not promote reactions with the electrolyte and are a promising candidate as anode materials.

We also calculated the average charge donation of Li/Na in the covalent crystals. For the configurations shown in Fig. 5 and 6, up to 0.69/0.81 |$e$| (Mulliken) and 0.32/0.33 |$e$| (Voronoi) per Li/Na atom is donated to TCNE units.

During the insertion of Li/Na, similar (Li/Na)$_x$-TCNE stoichiometries may be formed with both vdW and covalent crystals. For example, the insertion of one alkali atom per TCNE unit into the vdW crystal results in the same stoichiometry as that of the pristine covalent Li/Na-TCNE crystal. The total energy (of the lowest energy configuration) is about 0.1/0.3 eV (per TCNE unit) lower than the energy of the covalent Li/Na-TCNE crystal. At higher Li concentrations, structures obtained with the vdW crystal become higher energy (by 0.1 eV at Li$_2$-TNE and by about 0.7 eV at Li$_3$-TCNE) than those obtained with the covalent crystal at the same stoichiometry. At higher Na concentrations, structures obtained with the vdW crystal remain lower in energy than those obtained from the covalent crystal at the same stoichiometry (by 0.1 eV at Na$_2$-TCNE and by 0.2 eV at Na$_3$-TCNE). However, the geometries of the two types of crystals remain significantly different. This, together with the facts that (i) the Li-TCNE covalent structure was stable in Ref. [27] even though it is computed to be slightly higher in energy and (ii) there is a high energy cost to remove Li/Na from their positions in the pristine ecovalent crystals, suggests that the two types of crystals might not interconvert, although a definitive answer should be given by an experiment.

4. Conclusions

We carried out a comparative dispersion-corrected density functional theory (DFT) study of the possibilities of Li and Na storage in tetracyanoethylene(TCNE)-based vdW (molecular) and covalent crystals. Firstly, we confirmed the structure of previously synthesized (covalent) Li-TCNE crystal as well as predicted the existence of its Na-TCNE analogue. Then for both kinds of crystals, after a search of insertion sites, including insertion of multiple atoms, we find that up to 5.5/3 and 2.5/2 Li/Na atoms per TCNE unit can be inserted in the molecular and covalent crystals, respectively, while preserving the structure. The computed voltage can reach 3.54/2.16 V vs. Li/Li$^+$ for Li insertion and 3.31/2.67 V vs. Na/Na$^+$ for Na insertion in the molecular and covalent crystals respectively. Significantly, up to capacity of 418 mAh g$^{-1}$ for both Li and Na in the molecular crystal and 198 mAh g$^{-1}$ for Li and 177 mAh g$^{-1}$ for Na in the covalent crystal, the insertion of Li and Na would not lead to reactions with the electrolyte. This is a desirable property for both conventional liquid carbonate based electrolytes as well as for some of the promising solid state electrolytes (SSE) such LGPS (Li$_{10}$GeP$_2$S$_{12}$) and other sulfide SSE that do not form a self-arresting solid electrolyte interface [50]. Therefore, we conclude that tetracyanoethylene (TCNE) - based molecular and covalent crystals could become an efficient organic cathode and anode material, respectively, for both Li and Na ion batteries.

## 5. Acknowledgements

This work was supported by Tier 2 AcRF grant MOE2014-T2-2-006 by the Ministry of Education of Singapore.

## 6. References

[1] E. Barbour, I.A.G. Wilson, I.G. Bryden, P.G. McGregor, P.A. Mulheran, P.J. Hall, Towards an objective method to compare energy storage technologies: development and validation of a model to determine the upper boundary of revenue available from electrical price Arbitrage, *Energy Environ. Sci.*, **5** (2012) 5425-5436.
[2] D. Yue, P. Khatav, F. You, S.B. Darling, Deciphering the uncertainties in life cycle energy and environmental analysis of organic photovoltaics, *Energy Environ. Sci.*, **5** (2012) 9163-9172.
[3] T. Kousksou, P. Bruel, A. Jamil, T. El Rhafiki, Y. Zeraouli, Energy storage: applications and challenges, *Sol. Energy Mater. Sol. Cells*, **120, Part A** (2014) 59-80.

[4] M.M. Thackeray, C. Wolverton, E.D. Isaacs, Electrical energy storage for transportation-approaching the limits of, and going beyond, lithium-ion batteries, *Energy Environ. Sci.*, **5** (2012) 7854-7863.

[5] J.B. Goodenough, Electrochemical energy storage in a sustainable modern society, *Energy Environ. Sci.*, **7** (2014) 14-18.

[6] C.J. Barnhart, M. Dale, A.R. Brandt, S.M. Benson, The energetic implications of curtailing versus storing solar- and wind-generated electricity, *Energy Environ. Sci.*, **6** (2013) 2804-2810.

[7] C. Lupi, M. Pasquali, Electrolytic nickel recovery from lithium-ion batteries, *Miner. Eng.*, **16** (2003) 537-542.

[8] V. Palomares, P. Serras, I. Villaluenga, K.B. Hueso, J. Carretero-Gonzalez, T. Rojo, Na-ion batteries, recent advances and present challenges to become low cost energy storage systems, *Energy Environ. Sci.*, **5** (2012) 5884-5901.

[9] J.M. Tarascon, Is lithium the new gold?, *Nature Chem.*, **2** (2010) 510-510.

[10] M.D. Slater, D. Kim, E. Lee, C.S. Johnson, Sodium-Ion Batteries, *Adv. Funct. Mater.*, **23** (2013) 947-958.

[11] O.I. Malyi, T.L. Tan, S. Manzhos, A comparative computational study of structures, diffusion, and dopant interactions between Li and Na insertion into Si, *Appl. Phys. Express*, **6** (2013).

[12] O.I. Malyi, T.L. Tan, S. Manzhos, In search of high performance anode materials for Mg batteries: computational studies of Mg in Ge, Si, and Sn, *J. Power Sources*, **233** (2013) 341-345.

[13] O. Malyi, V.V. Kulish, T.L. Tan, S. Manzhos, A computational study of the insertion of Li, Na, and Mg atoms into Si(111) nanosheets, *Nano Energy*, **2** (2013) 1149-1157.

[14] F. Legrain, O.I. Malyi, S. Manzhos, Comparative computational study of the diffusion of Li, Na, and Mg in silicon including the effect of vibrations, *Solid State Ionics*, **253** (2013) 157-163.

[15] Y.L. Liang, Z.L. Tao, J. Chen, Organic Electrode Materials for Rechargeable Lithium Batteries, *Adv. Energy Mater.*, **2** (2012) 742-769.

[16] Z. Song, H. Zhou, Towards sustainable and versatile energy storage devices: an overview of organic electrode materials, *Energy Environ. Sci.*, **6** (2013) 2280-2301.

[17] H. Chen, M. Armand, G. Demailly, F. Dolhem, P. Poizot, J.-M. Tarascon, From Biomass to a Renewable LiXC6O6 Organic Electrode for Sustainable Li-Ion Batteries, *ChemSusChem*, **1** (2008) 348-355.

[18] A. Abouimrane, W. Weng, H. Eltayeb, Y. Cui, J. Niklas, O. Poluektov, K. Amine, Sodium insertion in carboxylate based materials and their application in 3.6 V full sodium cells, *Energy Environ. Sci.*, **5** (2012) 9632-9638.

[19] Y. NuLi, Z. Guo, H. Liu, J. Yang, A new class of cathode materials for rechargeable magnesium batteries: organosulfur compounds based on sulfur–sulfur bonds, *Electrochem. Commun.*, **9** (2007) 1913-1917.

[20] A.K. Shukla, T.P. Kumar, Lithium economy: will it get the electric traction?, *J. Phys. Chem. Lett.*, **4** (2013) 551-555.

[21] Y. Hanyu, I. Honma, Rechargeable quasi-solid state lithium battery with organic crystalline cathode, *Scientific Reports*, **2** (2012).

[22] Y. Park, D.S. Shin, S.H. Woo, N.S. Choi, K.H. Shin, S.M. Oh, K.T. Lee, S.Y. Hong, Sodium Terephthalate as an Organic Anode Material for Sodium Ion Batteries, *Adv. Mater. (Weinheim, Ger.)*, **24** (2012) 3562-3567.

[23] T. Yasuda, N. Ogihara, Reformation of organic dicarboxylate electrode materials for rechargeable batteries by molecular self-assembly, *Chem. Commun. (Cambridge, U. K.)*, **50** (2014) 11565-11567.

[24] N. Ogihara, T. Yasuda, Y. Kishida, T. Ohsuna, K. Miyamoto, N. Ohba, Organic dicarboxylate negative electrode materials with remarkably small strain for high-voltage bipolar batteries, *Angewandte Chemie - International Edition*, **53** (2014) 11467-11472.

[25] N. Dardenne, X. Blase, G. Hautier, J.-C. Charlier, G.-M. Rignanese, Ab Initio Calculations of Open-Cell Voltage in Li-Ion Organic Radical Batteries, *J. Phys. Chem, C*, **119** (2015) 23373-23378.

[26] Y. Chen, S. Manzhos, Lithium and sodium storage on tetracyanoethylene (TCNE) and TCNE-(doped)-graphene complexes: A computational study, *Mater. Chem. Phys.*, **156** (2015) 180-187.

[27] J.H. Her, P.W. Stephens, R.A. Davidson, K.S. Min, J.D. Bagnato, K. van Schooten, C. Boehme, J.S. Miller, Weak ferromagnetic ordering of the Li+ TCNE (center dot-) (TCNE = tetracyanoethylene) organic magnet with an interpenetrating diamondoid structure, *J. Am. Chem. Soc.*, **135** (2013) 18060-18063.

[28] Y. Chen, S. Manzhos, Li and Na storage on TCNE: a computational study, in: Proceedings of the 14th Asian Conference on Solid State Ionics (ACSSI-2014), Research Publishing Services, Singapore, pp. 374-380.

[29] Y. Chen, S. Manzhos, Li Storage on TCNE and TCNE-(Doped)-Graphene Complexes: a Computational Study, *MRS Online Proceedings Library*, **1679** (2014) DOI: 10.1557/opl.2014.1849.

[30] M. Khazaei, M. Arai, T. Sasaki, Y. Kawazoe, Polymerization of Tetracyanoethylene under Pressure, *J. Phys. Chem, C*, **117** (2013) 712-720.

[31] W. Kohn, L.J. Sham, Self-consistent equations including exchange and correlation effects, *Phys. Rev.*, **140** (1965) A1133-A1138.

[32] M.S. José, A. Emilio, D.G. Julian, G. Alberto, J. Javier, O. Pablo, S.-P. Daniel, The SIESTA method for ab initio order- N materials simulation, *J. Phys.: Condens. Matter*, **14** (2002) 2745.

[33] J.P. Perdew, K. Burke, M. Ernzerhof, Generalized gradient approximation made simple, *Phys. Rev. Lett.*, **77** (1996) 3865-3868.

[34] C. Kittel, Introduction to Solid State Physics, Wiley, Hoboken, NJ, 2005.

[35] E. Kaxiras, Atomic and Electronic Structure of Solids, Cambridge University Press, Cambridge, 2003.

[36] D. Bharaniaa, N. Capel, S. Manzhos, A comparative Density Functional Theory and Density Functional Tight Binding Study of interfaces of a high energy density material N8, in.

[37] H.J. Monkhorst, J.D. Pack, Special points for brillouin-zone integrations, *Phys. Rev. B*, **13** (1976) 5188-5192.

[38] C. Fonseca Guerra, J.-W. Handgraaf, E.J. Baerends, F.M. Bickelhaupt, Voronoi deformation density (VDD) charges: assessment of the Mulliken, Bader, Hirshfeld, Weinhold, and VDD methods for charge analysis, *J. Comput. Chem.*, **25** (2004) 189-210.

[39] S. Grimme, Semiempirical GGA-type fensity functional constructed with a long-range dispersion correction, *J. Comput. Chem.*, **27** (2006) 1787-1799.

[40] W. Luo, J. Wan, B. Ozdemir, W. Bao, Y. Chen, J. Dai, H. Lin, Y. Xu, F. Gu, V. Barone, L. Hu, Potassium Ion Batteries with Graphitic Materials, *Nano Lett.*, **15** (2015) 7671-7677.

[41] G. Ceder, G. Hautier, A. Jain, S.P. Ong, Recharging lithium battery research with first-principles methods, *MRS Bulletin*, **36** (2011) 185-191.

[42] R. Mukhopadhyay, S.L. Chaplot, Structural changes in tetracyanoethylene at high pressures: neutron diffraction study, *J. Phys.: Condens. Matter*, **14** (2002) 759.

[43] D.A. Bekoe, K.N. Trueblood, *Z. Kristallogr.*, **113** (1960) 1-22.

[44] H. Yamawaki, K. Aoki, Y. Kakudate, M. Yoshida, S. Usuba, S. Fujiwara, Infrared study of phase transition and chemical reaction in tetracyanoethylene under high pressure, *Chem. Phys. Lett.*, **198** (1992) 183-187.

[45] R.G. Little, D. Pautler, P. Coppens, X-ray structure analysis of cubic tetracyanoethylene and the length of the C[triple bond] N bond. Application of a double-atom refinement method, *Acta Crystallographica Section B*, **27** (1971) 1493-1499.

[46] N. Capel, D. Bharania, S. Manzhos, A Comparative Density Functional Theory and Density Functional Tight Binding Study of Phases of Nitrogen Including a High Energy Density Material N8, *Computation*, **3** (2015) 574.

[47] J.B. Goodenough, Y. Kim, Challenges for rechargeable Li batteries†, *Chem. Mater.*, **22** (2009) 587-603.

[48] S. Komaba, W. Murata, T. Ishikawa, N. Yabuuchi, T. Ozeki, T. Nakayama, A. Ogata, K. Gotoh, K. Fujiwara, Electrochemical Na insertion and solid electrolyte interphase for hard-carbon electrodes and application to Na-ion batteries, *Adv. Funct. Mater.*, **21** (2011) 3859-3867.

[49] A. Bhide, J. Hofmann, A. Katharina Durr, J. Janek, P. Adelhelm, Electrochemical stability of non-aqueous electrolytes for sodium-ion batteries and their compatibility with Na0.7CoO2, *Phys. Chem. Chem. Phys.*, **16** (2014) 1987-1998.

[50] A.C. Luntz, J. Voss, K. Reuter, Interfacial Challenges in Solid-State Li Ion Batteries, *J. Phys. Chem. Lett.*, **6** (2015) 4599-4604.

[51] K. Momma, F. Izumi, VESTA 3 for three-dimensional visualization of crystal, volumetric and morphology data, *J. Appl. Crystallogr.*, **44** (2011) 1272-1276.

**Figures**

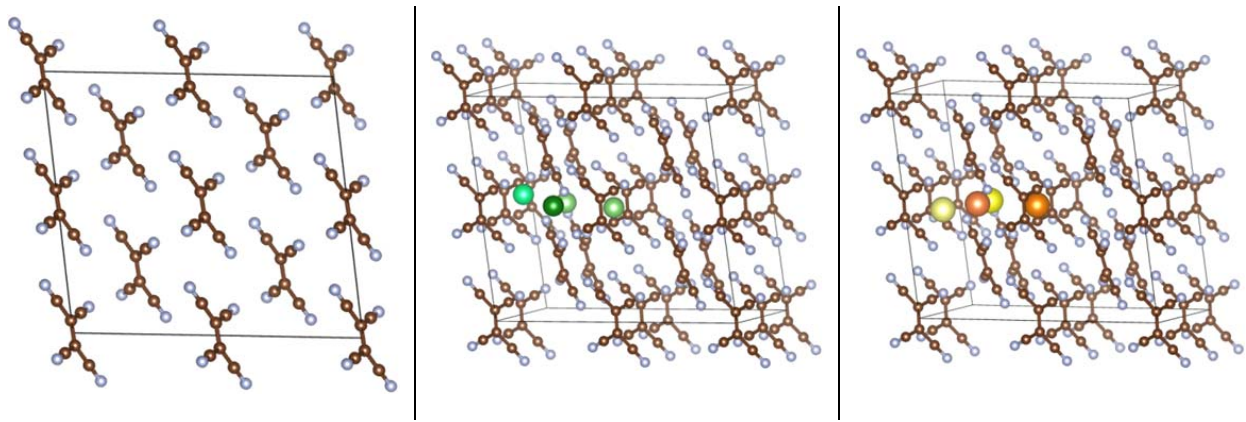

Figure 1. The crystal structure of monoclinic TCNE (left) and the insertion sites of Li (middle) and Na (right) atoms in the crystal. Atom colour scheme here and elsewhere: C-brown, N-grey, Li-green and Na-yellow. Different shades of green and yellow are used for different insertion sites. Visualization here and elsewhere is by VESTA [51].

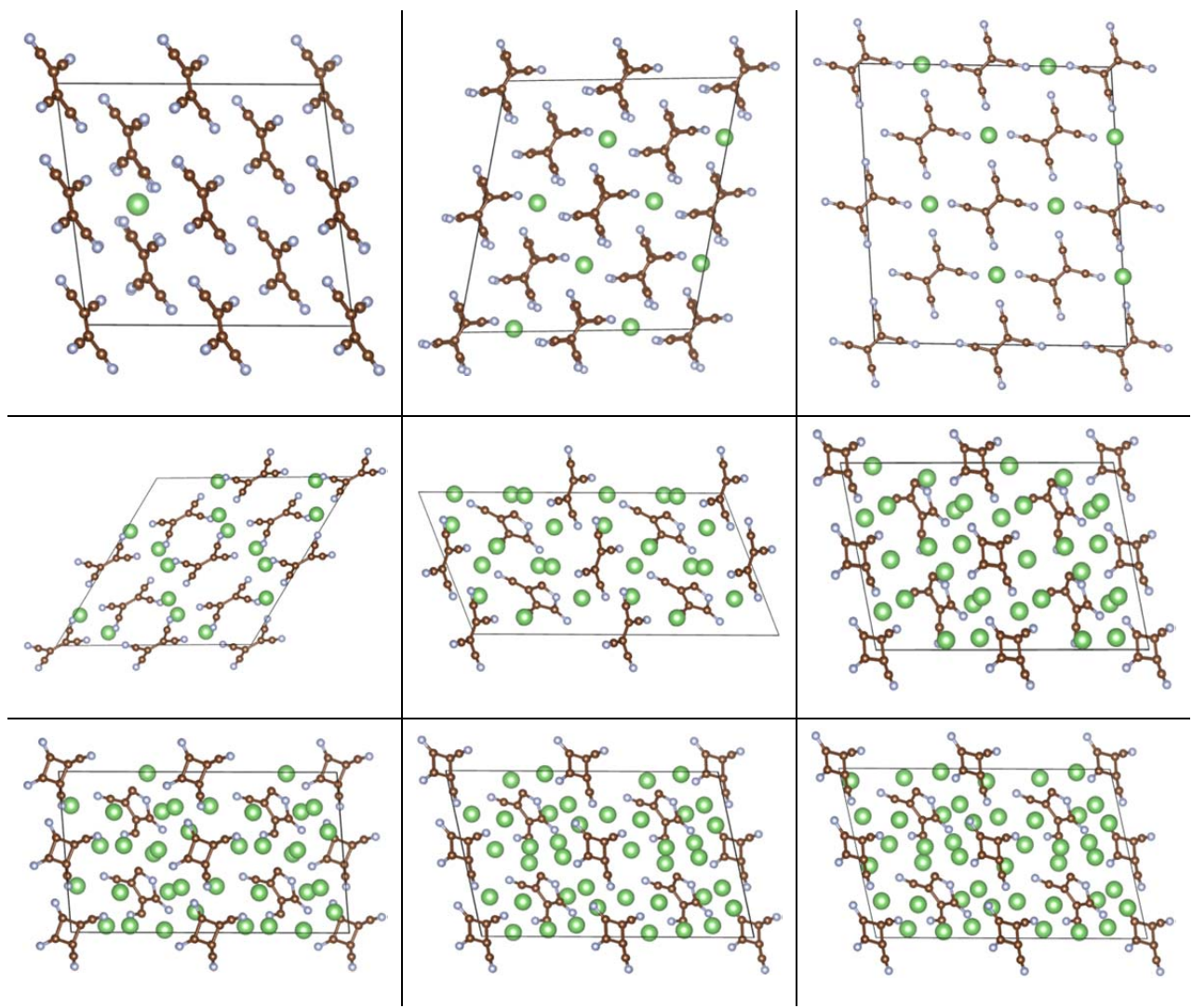

Figure 2. Configurations of Li$_m$-TCNE with the strongest $E_b$, left to right and top to bottom: $m$=0.0625, 0.5, 1, 2, 3, 4, 4.5, 5, 5.5.

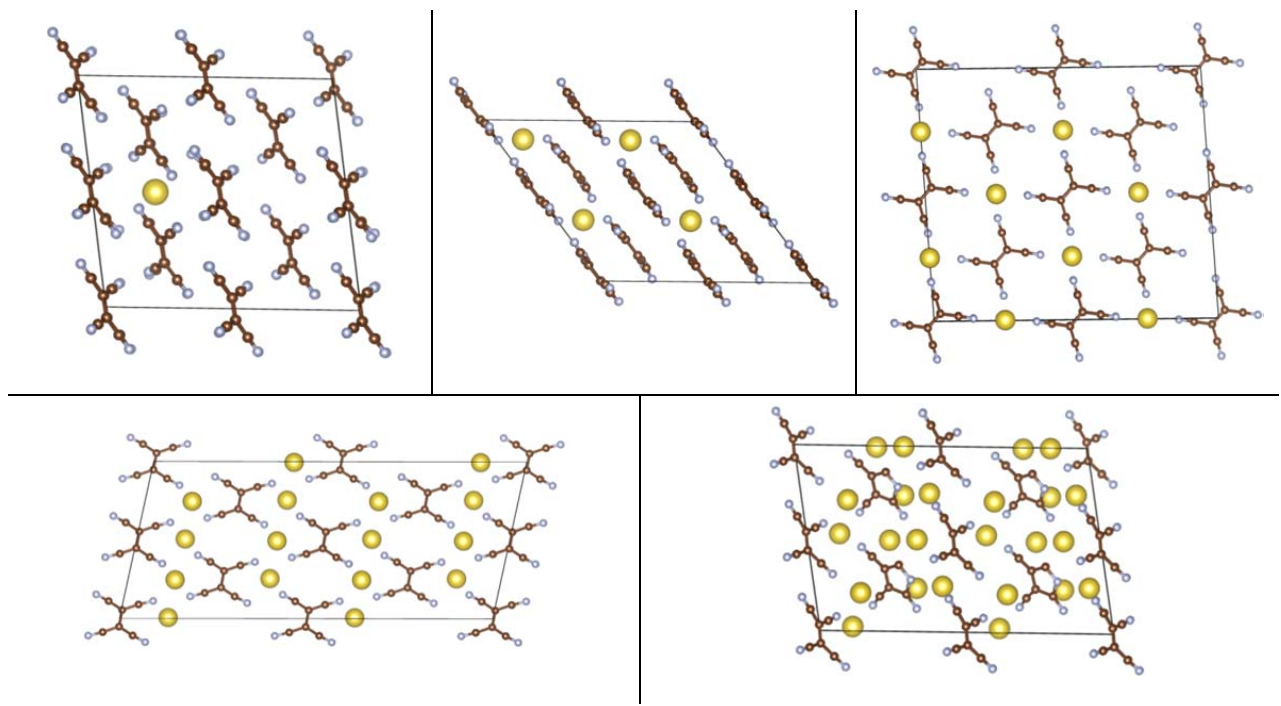

Figure 3. Configurations of Na$_n$-TCNE crystals with the strongest $E_b$, left to right and top to bottom: $n$=0.0625, 0.5, 1, 2, 3.

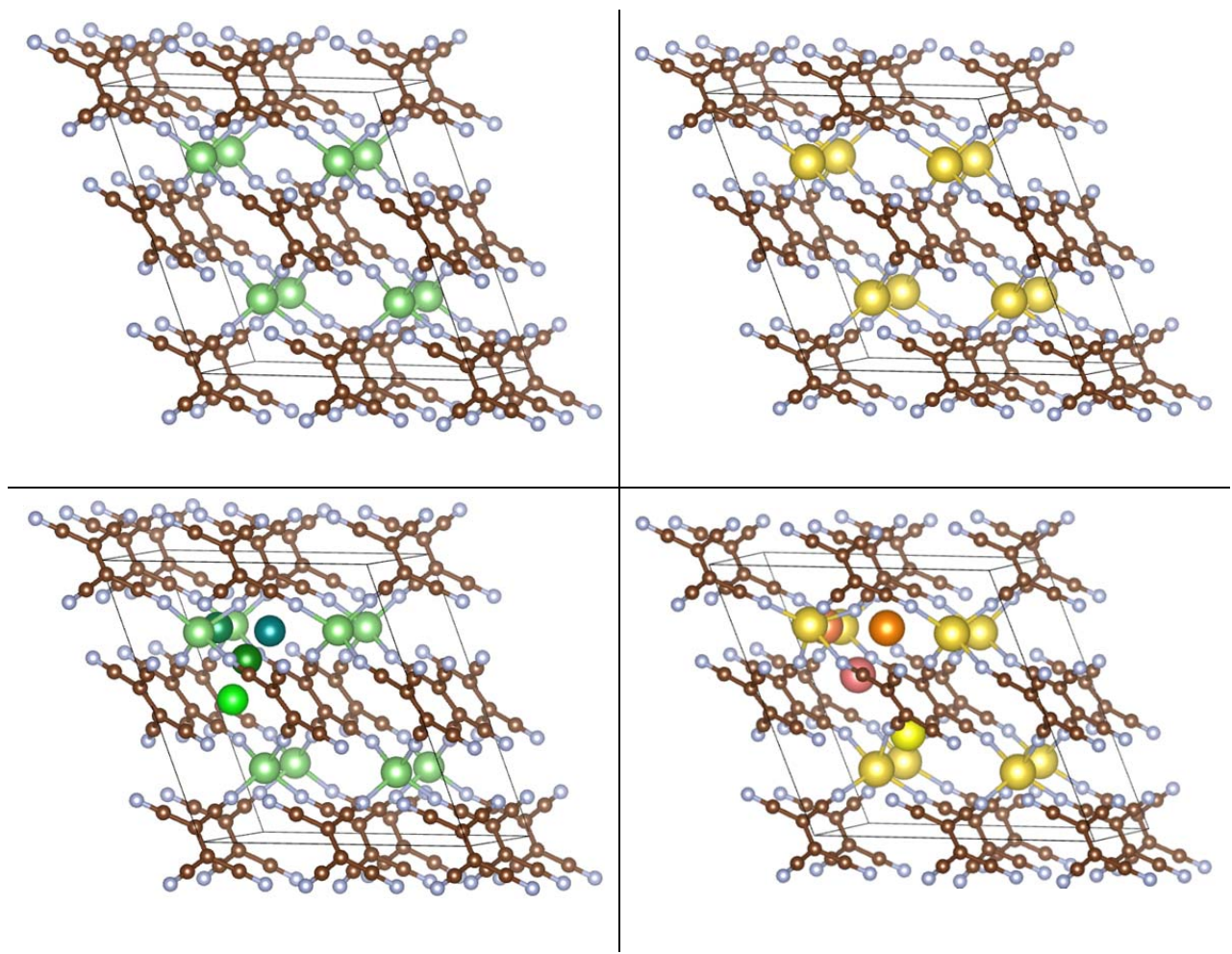

Figure 4. Top panels: the crystal structures of covalent Li-TCNE (left) and Na-TCNE (right) crystals. Bottom panels: insertion sites of Li (left) and Na (right) atoms in the crystals (different shades of green/yellow are used for different Li/Na sites).

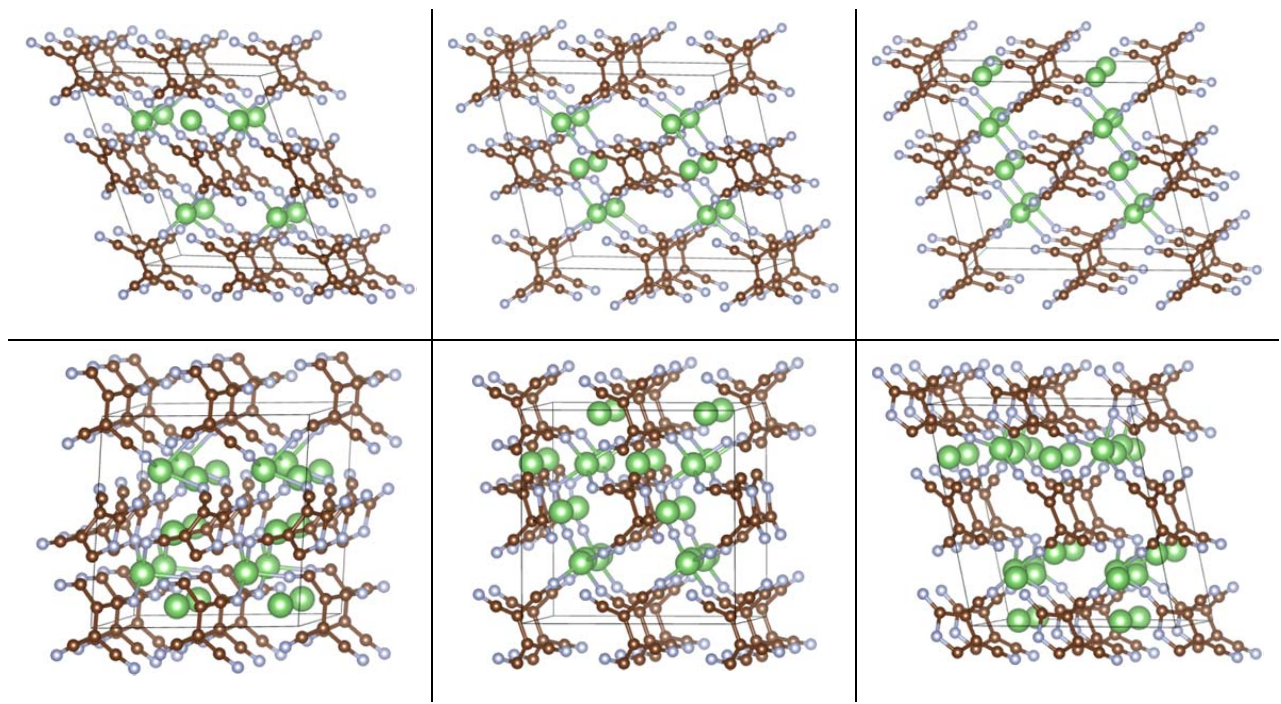

Figure 5. Configurations of Li$_m$-(Li-TCNE) crystals with the strongest $E_b$, left to right and top to bottom: $m$=0.125, 0.5, 1, 1.5, 2, 2.5.

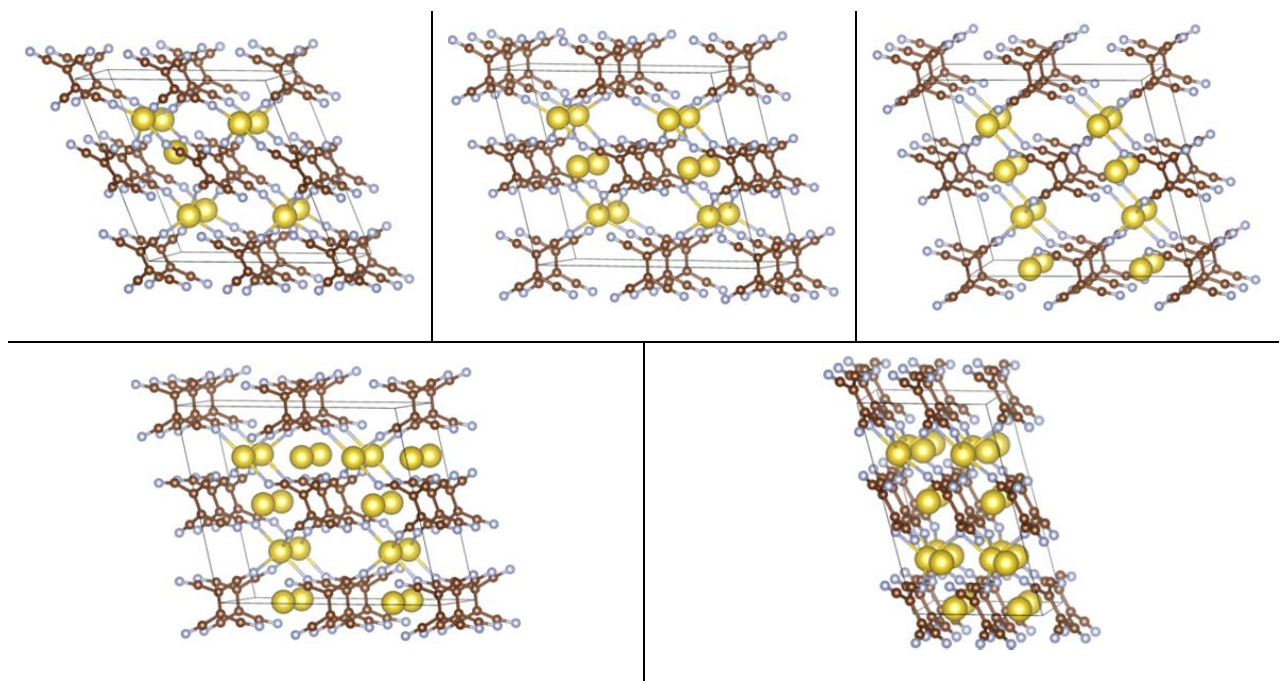

Figure 6. Configurations of Na$_n$-(Na-TCNE) crystals with the strongest $E_b$, left to right and top to bottom $n$=0.125, 0.5, 1, 1.5, 2.

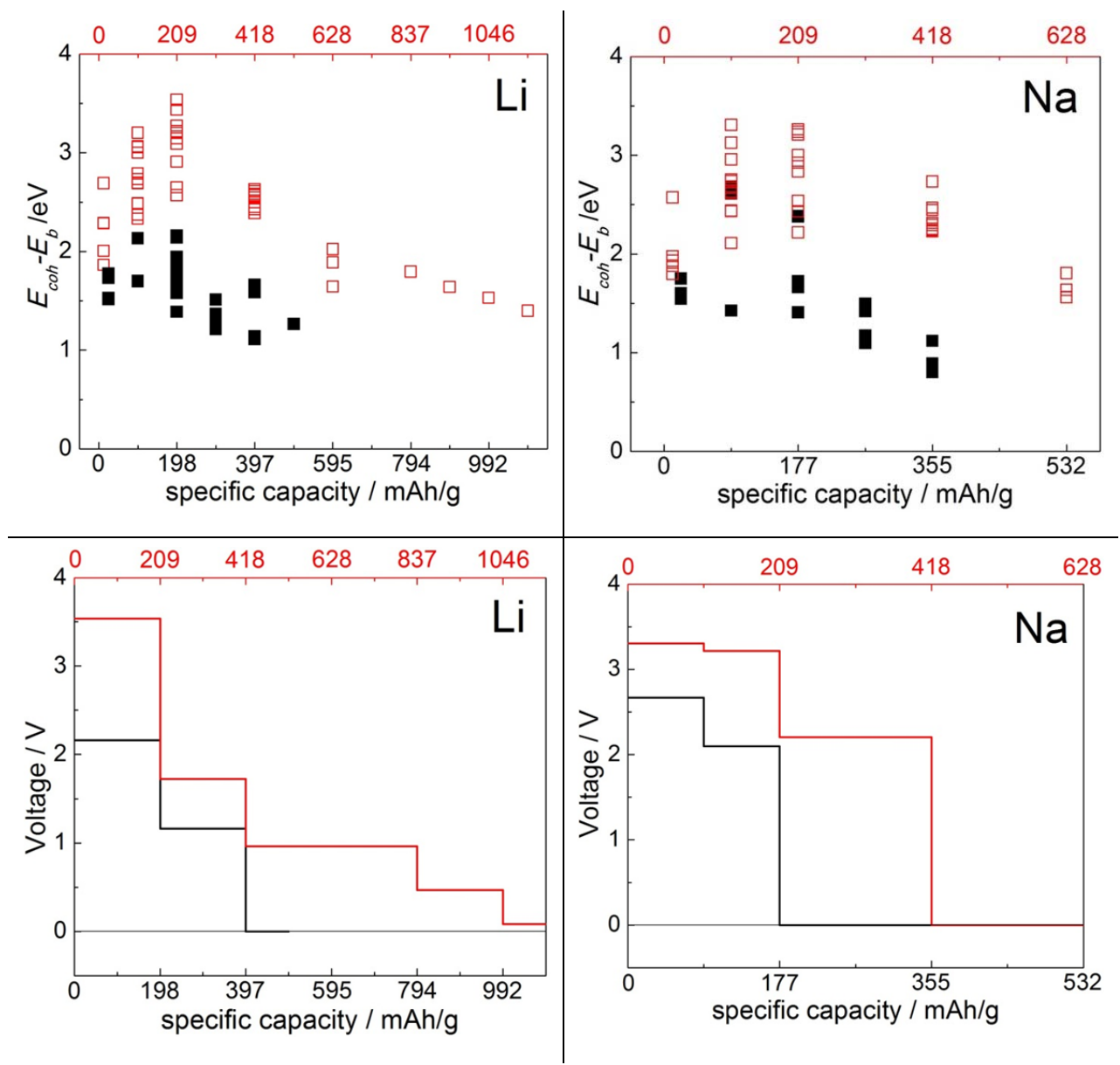

Figure 7. Top: binding strength per alkali atom on the "voltage" scale ($E_{coh}^{Li/Na}-E_b$) vs. specific capacity corresponding to the no. of inserted Li/Na atoms in the monoclinic TCNE crystal (red empty squares and top axes) and covalent Li/Na-TCNE crystal (black filled squares and bottom axes). Bottom: computed voltage profiles vs. specific capacity in the monoclinic TCNE crystal (red lines and top axes) and covalent Li/Na-TCNE crystal (black lines and bottom axes).